Rolling Release Siege Engines: Teaching and Old Machine a New Trick

Dr. Joseph O. West[a)], Seth Ross, and James Flesher

Department of Chemistry and Physics

Indiana State University

Terre Haute, IN 47809

**Abstract**

The analysis of a new "rolling release" mechanism is presented for three different "siege engine" designs. The range obtained with the rolling release is compared to the ranges obtained using the simpler "cup" and the proven "sling" release mechanisms. It is found that the rolling release is a significant improvement over the cup release, but the range of the rolling release is still well short of that attained with the sling release. It is also noted that the familiar "spoon" projectile holder used on many siege engines operates in a rolling release manner, contrary to appearances and to some previous published studies. Instead, the release mechanism used in ancient engines and in modern kit-purchased replicas is the rolling release.

## I. INTRODUCTION

Siege engines hold a certain fascination with history buffs, engineers and physicists.[1] There are annual siege engine competitions held throughout the nation in high schools, college physics and engineering departments, and even some "open to the public" competitions at the national level. A wide range of kits are available commercially for building siege engine replicas with the completed models ranging in size from those small enough to fit on a typical watch face, to those large enough to throw bowling balls more than 100 meters. The mathematical models of this class of machines are well developed. Computer programs are available for download, some for free, that are meant to be used to evaluate various design parameters.[2,3] Siege engines have been the topic of documentaries,[4] featured on multiple Mythbusters episodes,[5,6] and have made appearances in fantasy feature films.[7] Investigators are cautioned to keep safety concerns in mind when working with the reconstructions, as siege engines were originally designed as weapons of war. Some working models have been large enough to launch people, at least once with tragic and deadly consequences.[8]

In this paper, a rolling release, which is a "new" variation of an "old" release method, is analyzed. The new method improves the range of siege engines over that of a simple "cup" release. It is the opinion of the authors that the classic torsion powered "onager," with its easily recognized "spoon" projectile holder [Fig. 1] has actually always used the rolling release. One uses a variation of this design to launch peas from a plastic spoon at meal time. As observed in a study of the performance of recreations of ancient weapons in the late 1800s, the projectile actually leaves the spoon well before the spar makes contact with the crossbeam.[9] This is contrary to the common assumption that the projectile is launched with a velocity perpendicular to the spar.[10] The "new" rolling release idea was tested in a kit constructed by the authors and applied in a mathematical model in which the ball is allowed to roll along the entire length of the spar. The initial position of the ball on the spar is determined by optimizing the range of the engine. This allows more flexibility in the design than the spoon holder, and greatly decreases the complexity of the mathematics needed to describe the dynamics. Because the projectile rolls to the end of the spar, it leaves the engine with a known and significant angular velocity ("backspin"). The aerodynamic lift on the ball described by the Magnus effect is included in addition to the usual wind resistance forces. The ranges attained by the engines employing the cup, rolling, and sling release method are determined for a suitable set of design parameters. The paper focuses on gravity powered engines, but results for torsion powered variants are also included. It is found for both gravity and torsion powered engines that the rolling release is superior to the cup release, but it falls short of the performance of the well established sling release. The physics of the cup and rolling releases is suitable for short projects for advanced undergraduates at the junior and senior level. The various machine designs are described in Section II, the details of the calculations are presented in Section III, the comparison of the ranges is presented in Section IV, and conclusions are presented in Section V.

## II. THE MACHINE DESIGNS

Four counterweight gravity powered siege engine designs are considered: Tosser (cup release), Slinger (sling release), Roller (rolling release), and Hinged Roller (rolling release). In operation, a massive counterweight is mounted on the main spar, and the crew does work rotating the spar, usually from vertical to an angle beyond horizontal, lifting the counterweight. When loosed, the gravitational potential energy of the counterweight converted to kinetic energy much like an unbalanced teeter-totter. A very complete analysis of the efficiency and range comparisons of the Tosser and Slinger has recently been completed by Denny,[11] and the reader is directed to that work for much of the details and the equations of motion of those designs. In the Hinged Roller, the counterweight is attached to the spar with a hinge so that the counterweight is allowed to swing freely when the machine is loosed.

All of the calculations were performed assuming that the projectile is a uniform sphere, i.e. a ball. Calculations of the trajectory of the ball included wind resistance, and when launched with the rolling release, the Magnus effect. The Magnus effect describes a force perpendicular to the direction of motion, and for a ball like those considered here with "backspin," has a positive vertical component throughout the flight. The effects of this force can be demonstrated in a dramatic fashion with the widely available and inexpensive plastic ball throwing toy, the Wham-O Trac ball®.[12] That toy uses a curved plastic basket that looks very much like the cesta of jai-alai to throw a hollow plastic ball with a diameter of approximately 10 cm. The ball rolls along the plastic launcher and becomes a projectile when it reaches the end of the basket. Even when the ball is initially projected horizontally, it is possible to get the ball to **gain** altitude for a short time. A similar device has been described for demonstrating the Bernoulli effect in the classroom with a ping pong ball.[13] In the projectile phase of its motion, the forces on the ball are then[14-17]

$$F_{wx} = -Cv^2 \left( \frac{2v_x + v_y}{2v} \right)$$
$$F_{wy} = -Cv^2 \left( \frac{v_x - 2v_y}{2v} \right) - mg, \qquad (1)$$
$$C = 0.22(\text{diameter})^2$$

where v is the speed of the ball relative to the spar at the moment it leaves the spar; $v_x$ and $v_y$ are the x and y components of the velocity of the ball in the air; m is the mass of the ball, and g is 9.8

m/s$^2$. When the ball is loosed from the Tosser and Slinger minimal spin is expected and only the standard quadratic wind resistance forces and the force of gravity are included.[18]

## IIa. Tosser

The Tosser is a "teeter-totter" equipped with a simple cup to hold the ball, as shown in Fig. 2. This is the "basic lever arm engine" of Denny.[11] The launch angle θT is the angle of the spar above horizontal at which the spar collides with the crossbeam. Here the optimal value of θT is determined numerically by maximizing the range with respect to θT. The angular velocity of the spar can be determined analytically from conservation of energy,[10,11,19] and if one neglects wind resistance the range can also be determined analytically. However, the range was calculated numerically, in order to keep the procedures identical for all of the engine models.

## IIb. Slinger

In ancient times, a major advance in siege engine designs was the addition of a sling to the end of the spar. Such a sling turns the Tosser into the Slinger, as shown in Fig. 3. The Slinger is referred to as the "mangonel: fixed counterweight" by Denny.[11] Here the optimal value of SL is determined numerically by maximizing the range with respect to SL. Consistent with Denny,[11] it is assumed that the sling has been "tuned" so that the ball is launched when the **sling** is at 45 degrees relative to the ground, and the ball is projected perpendicular to its length, regardless of the angle of the spar. The sling is assumed initially to lie flat on the ground, and to move freely starting from the launch. In actual operation, the ball slides along the ground for a short distance, but this sliding phase is expected to have only a minor influence on the overall dynamics of the machine, and is not included in the calculations.[11] The addition of the sling effectively increases the length of the spar, but with very minimal increase in the mass of the spar.[10,11] The sling and "pouch" holding the ball are assumed to be of negligible mass. There is also a "whip" action of the sling, such that the angular velocity of the sling at the time of launch is much greater (factor of two or so) than that of the spar. The resulting dynamics are very efficient at transferring energy to the ball, and are similar to the dynamics of the "two segment model" of the throwing motion of the human arm used by Cross.[20] The primary safety issue for the operator and bystanders in the case of the Slinger is the metal hook used to attach the sling to the end of the spar. The spar and hook accelerate to high speed over a short time interval when loosed.

**IIc. Roller**

The Roller provides a more accurate model than the Tosser for the release used in the onager. Sir Ralph Payne-Gallwey, in his classic examination of crossbows, ballista and catapults reported on the performance of a replica of a torsion powered spoon design and specifically noted that the ball leaves the spoon **before** the spar impacts with the crossbeam.[9] The same mechanics are utilized in the sports of lacrosse (the stick with basket is equivalent to a spoon) which produces moderate ball launch speeds, and in jai-alia (the cesta is more like the spar) in which the ball can reach speeds of 180 mph.[21] As noted by Denny,[10] the optimal angle for the spar at launch for the Tosser is **not 90 degrees**, and yet that is the mathematical model that immediately comes to mind when one first sees the onager crossbeam set at 90 degrees. The authors are not aware of any historical or commercially available design that utilizes a cross beam short of 90 degrees, contrary to what one would expect from the results of the analysis of the Tosser.

In the Roller, the spoon is removed from the onager and the ball is simply allowed to roll along the top of the spar [Fig. 4]. The kit the authors used to test this release model was modified by the addition of guide rails to the spar to keep the ball from moving laterally off of the spar during launch [Fig. 5]. The removal of the spoon greatly simplifies the mathematics, and allows greater freedom in optimizing the range. The authors developed a model of the "spoon" release, but the resulting equations of motion for even the simplest choice of spoon geometry proved to be too complicated for practical use. The equations of motion for the Roller are then

$$\ddot{R} = \left(\frac{5}{7}\right)\left(R\dot{\theta}^2 + g\sin\theta\right)$$
$$\ddot{\theta} = \left(\frac{MD - M_{spar}(L-D) - mR}{MD^2 + I + mR^2}\right)g\cos\theta - \frac{2mR\dot{R}\dot{\theta}}{MD^2 + I + mR^2}, \qquad (2)$$

where $\theta$ is the angle the spar makes with respect to horizontal; R is the distance from the pivot to the location of the ball on the spar; $M_{spar}$ is the mass of the spar; I is the moment of inertia of the spar; D is the distance of the counterweight from the pivot; M is the mass of the counterweight; and L is the distance from the pivot to the launch end of the spar. The total length of the spar is L + D, and the value I is found to be: $I = (1/3)M_{spar}(L^2 + D^2)$. The ball becomes airborne when it reaches the end of the spar, i.e. when R = L. The ball leaves the spar with significant "radial" (dR/dt) **and** tangential (Rd$\theta$/dt) velocity components so the launch speed is **greater** than the tip of the spar. This is the primary advantage of the Roller over the Tosser. The angle of the spar at launch, and the range, are both determined by the initial location of the ball on the spar, RL. Here the optimal value of RL is determined numerically by maximizing the range with respect to RL. The Roller allows the use of a variety of projectiles, including "bolts" which could **slide** off the end of the spar. The equations of motion for sliding projectiles are obtained from Eq. (2) by replacing the factor of 5/7 with 1. The range of the sliding bolt (ink pen) used for testing was only slightly shorter than that of the ball. The bolt even arrived at the target "sharp" end first in the limited tests conducted. The cap should be glued in place on the ink pen before allowing students to use them as projectiles.

Porter and Tremblay presented a theoretical study of a "conceptual" siege engine very similar to the Roller.[19] In that work, it is assumed that the ball becomes airborne when the spar reaches 90 degrees, similar to the spar hitting a crossbeam in the Tosser. The speed of the ball at the end of the spar was determined using conservation of energy. It was **assumed** that the projectile would be launched at the angle that maximized the range, rather than horizontally. No specific suggestion as to why or how that particular release angle would be achieved was provided, and the derivation implicitly assumed that the projectile does not roll on the spar.

**IId. Hinged Roller**

An inherent drawback to the Roller is that the ball can not have a large tangential velocity component, unless the spar and the counterweight have a large angular velocity (and kinetic energy). It is already well know that the range of the Slinger can be increased by attaching the counterweight on a hinge, so that it is free to swing in the same manner as the sling, and independent of the main spar.[11] As an effort to make the Roller more competitive with the Slinger, a hinged counterweight is added to the Roller, transforming it into the Hinged Roller. A diagram of a Hinged Roller is shown in Fig. 5a, while a photograph of a commercial trebuchet kit constructed by the authors and modified by the removal of the sling and the addition of guiding rails is shown in Fig. 5b. In this design, the spar still attains a significant angular velocity, but the moment of inertia of the spar alone is much less than the moment of inertia of spar and counterweight combined, so that there is a large decrease in their rotational kinetic energy of the spar "assembly" when in the hinged configuration. The results of calculations show that the use of the hinge reduces the speed of the counterweight to only 40% of its speed in the Roller at the time of launch. The ball therefore has access to more kinetic energy. The dynamics of the Hinged Roller are more complicated, and the range is to be maximized based on the choice of **three** adjustable parameters: the initial location of the ball on the spar RH, the initial angle of the hinged counterweight ψ, and the length of the counterweight portion of the spar that is devoted to the swinging arm length A [Fig. 5a]. As with the Roller, bolts can be used in the Hinged Roller [Fig. 5c].

**IIe. Practical Considerations**

In gravity powered engines, the spar is expected to be at a relatively large angle relative to the horizon for loading. A method to prevent the ball from simply rolling down the length of spar and onto the ground before the machine is loosed is a challenging design issue for the Roller and Hinged Roller. The ball needs to be held in place when loaded with a block of some sort, resulting in a two-stage firing sequence (release ball, release spar). In torsion powered engines, the ball can be kept in place more easily. A simple rope tied around the spar at the desired location will suffice, as the spar makes a relatively small (<10$^o$) angle with respect to horizontal when in the loaded position. In testing the kit constructed Roller, one of the authors (JW) wore full eye and body safety gear, and simply placed the ball (or bolt) on the spar by hand. The ball and the spar were then released "simultaneously" by hand, but only for a few trials. This is NOT suitable or safe as a regular launching method.

**III. COMPUTATIONAL METHODS**

The dynamics of the ball, the spar, and the hinged counterweight (Hinged Roller only) were simulated numerically using a standard fourth-order Runge-Kutta method, with a time step of 0.00001 seconds. This time step is excessively small, given the lack of precision in constructing real engines and in the approximations made in the mathematical models. The run times for the programs, even when optimizing the range of the Hinged Roller with three adjustable parameters, are all less than 30 seconds so that the time step is not a practical constraint. If one is wishing to animate the dynamics of the engines on a computer, the time step could be adjusted for computer speed to provide simulations that run in "real time." Two of the authors (JW and SR) each wrote a stand-alone program in Visual Basic for each engine considered, except the Hinged Roller. The outputs of these independently written programs were compared with each other to check for consistency. The Tosser and Slinger simulation programs were also tested by reproducing the results of Denny.[11] Only one Hinged Roller simulation program was written (JW), and it was written in the open source VPython language.[22] The code for the Roller design was modified to produce the Hinged Roller program, the required changes were minimal, and the resulting program was thoroughly tested.

In all cases considered, the same parameters were used: mass $m = 0.057$ kg; diameter $= 0.032$ m, consistent with a tennis ball; $D = 0.3$ m and $L = 0.96$ m; $M_{spar} = 0.559$ kg (modeled as a uniform beam of wood, approximately 1.0 inch square in cross section); and $M = 6.0$ kg. The values used are consistent with back yard and school competition trebuchet designs and those used in an analysis of a replica made to throw golf balls.[23] The pivot of the spar is modeled to be at a height of 0.2 m above the ground. The spar is loosed with the counterweight above ground level and the spar at an angle of 40 degrees. This value was chosen based on historical pictures, and the tests conducted on the kits constructed by the authors.

**IV. MACHINE EVALUATIONS**

The optimized ranges, and the associated parameters, are found in Table I. The range of the Roller is approximately twice that of the Tosser. However, the Slinger is clearly superior, with a range of roughly triple that of the Roller. The optimal initial position of the ball on the spar for the Roller is much less than the length of the spar (RL = 0.6L), suggesting that the range obtained using a spar with a spoon can be improved significantly by removing the spoon holder and instead allowing the ball to roll for a longer distance along the spar.

The introduction of the hinged counterweight, at the cost of two additional adjustable parameters, allows the Hinged Roller to roughly double the range of the more basic Roller. Notice that the optimal placement of the ball on the spar is even closer to the pivot than in the case of the Roller (RH = 0.3L = RL/2). The Hinged Roller still falls short of the performance of the Slinger, and there are additional "unresolved issues" with the Hinged Roller. The arm and attached counterweight have a significant angular velocity at the time of launch, and the **rotational** kinetic energy of the counterweight about its center of mass has **not** been included in the calculations. It is treated as a point mass in the simulations. Propping the hinged counterweight when loaded, but allowing it to swing freely upon launch is no small design challenge. The range of the Slinger can also be improved by the same hinged counterweight upgrade to become a Hinged Slinger so that the use of a sling remains advantageous, other design considerations held constant. The Hinged Slinger is more commonly referred to as a trebuchet.[11,23]

The results of calculations based on torsion powered versions of the machines (Torsion Tosser, Torsion Slinger and Torsion Roller) are also presented in Table I. In the torsion powered machines, the crew does work rotating the spar against the force of twisted ropes (skeins), from a nearly vertical position to a nearly horizontal position, storing potential energy in the ropes. The projectile (ball) is loaded, and when fired, the spar accelerates back to the vertical position, up to the point that it strikes (violently) the crossbeam. The torque due to the skeins was assumed to be linear in the angle of rotation, torque = - K$\theta$. In practice, skeins are under tension, even after the engine has been loosed, so that the angle of rotation was assumed to be $5\pi/2$ radians when the spar is in the loaded position. For comparison purposes, the torsion spring constant K, was chosen so that the net torque due to the coiled skeins and that due to the counterweight were the same with the spar held horizontal to the ground. The results obtained in comparing the three release methods were very similar to those found in the gravity powered models, and in the interest of brevity, the details of those models are not presented here.

In addition to the torsion powered variations, the gravity and torsion powered versions of the Roller and Hinged Roller were also modeled under the assumption of zero-friction "sliding release" of the projectiles such as the bolts mentioned previously. While the details of the engine dynamics are different, the ranges obtained using sliding and rolling projectiles are within 5% for a given engine. Due to the similarity in range, the details of those results are not presented here. It is somewhat difficult to justify an assumption of no friction for the sliding projectiles, given the large contact force between the spar and the projectile during the launch phase. However, the tests performed launching the pen from the Roller kit support the predicted small effect on the range. As a launcher of sliding projectiles (bolts or bunches of bolts) the Hinged Roller might still find use on the same battle field as the Slinger, despite the Hinged Roller's reduced range.

## V. CONCLUSIONS

The use of a rolling release in gravity powered siege engines has been investigated. The range of a machine using the rolling release is approximately twice the range obtained by a machine that utilizes the cup release, other design parameters held constant. The results indicate that the actual dynamics of the ball when released from a spoon holder in the engines of antiquity and contemporary kit designs is more accurately described by the rolling release, and not the cup release that is often assumed. The rolling of the ball along the spar increases the range of the engine, and this range can be optimized and controlled by the choice of initial position of the projectile on the spar. Sliding darts might also be viable projectiles in the same machine with minimal modifications in operating procedure. The use of a packet of darts might have been effective against personnel, while a single large dart might have been more useful against harder walls. Such a single large dart would provide a much more narrow point of impact than a ball. Consistent with the historical record, it is found that the use of a sling at the end of the spar clearly produces the greatest maximum range for both torsion and gravity powered siege engines.

The topic siege engines hold a certain fascination with engineers, physicists, and students. The mathematics of the cup release and rolling release models are accessible to undergraduates at the junior and senior level, while the construction of siege engines from scratch or from commercially available kits is accessible to a very wide audience. The engine parameters are apparent to the student, and the idea of optimizing the range of the engines has great appeal. Producing full animation of the dynamics of these engines using the VPython programming language would be a suitable project for a one semester junior level mechanics course, or a one semester computational physics course.

**ACKNOWLEDGEMENTS:**

The authors wish to thank Professor Kyle Lanoue and the Fall 2005 section of IMT 107 (Materials Engineering) at ISU for constructing two proof of concept gravity powered rolling release engines from 2x4 boards and free weights.

**REFERENCES**

a) Electronic mail: <joseph.west@insdtate.edu>

1) Play "destroy the castle" using a simulated trebuchet at:

<http://www.pbs.org/wgbh/nova/lostempires/trebuchet/destroy.html>.

2) D. B. Siano, website with information concerning siege engines in general, and a link to a program to estimate siege engine range based on the design parameters

<http://www.algobeautytreb.com/> .

3) See <www.RLT.com>.  They have links to just about everything for the enthusiast for sale.

4) Secrets of lost empires. Medieval siege, documentary produced by WGBH Video (2000, broadcast January 2006), South Burlington, VT color, available in VHS, 60 minutes.  "NOVA set two teams of timber framers, engineers, and historians the challenge of building precise replicas of this ultimate thirteenth century deterrent. Armed only with traditional tools…" <http://www.pbs.org/wgbh/nova/listseason/27.html#27ms>.  The companion web site: <http://www.pbs.org/wgbh/nova/lostempires/trebuchet/>.

5) MythBusters, Episode 22: Boom-Lift Catapult, produced by Discovery Communications, Silver Springs, MD, (aired November, 2004).

6) MythBusters, Episode 35: Border Slingshot, produced by Discovery Communications, Silver Springs, MD, (aired July, 2004).

7) Watch the siege of Minas Tirith in The Lord of the Rings: The Return of the King, film produced by New Line Cinema, Burbank, CA, color, released 2003 multiple formats, color 200 minutes.

8) Simon de Bruxelles, Lewis Smith and Alan Hamilton, "Oxford student killed in catapult stunt," The Times Online, From The Times, November 26, 2002 (accessed March 2010) http://www.timesonline.co.uk/tol/news/uk/article835651.ece.

9) Sir R. Payne-Gallwey, *The Crossbow* (Barns & Noble, USA 1996, originally published in 1903), p. 281.

10) M. Denny, "Optimum onager: the classical mechanics of a classical siege engine," Phys. Teach. **47**, 574 – 578 (2009).

11) M. Denny, "Siege engine dynamics," Eur. J. Phys. **26**, 561–577 (2005).

12) The Wham-O Trac ball®, information can be found at <http://www.wham-o.com/default.cfm?page=ViewProducts&ProductID=225&Category=3>.

13) R. E. Worely, "Bernoulli demonstration," Phys. Teach. **3**, 320 (1965).

14) M. K. McBeath, A. M. Nathan, A. T. Bahill, D. G. Baldwin, "Paradoxical pop-ups: Why are they difficult to catch?" Am. J. Phys. **76**, 723 – 729 (2008).

15) A. M. Nathan, "The effect of spin on the flight of a baseball," Am. J. Phys. **76**, 119- 124 (2008).

16) J. Rossmann and A. Rau, "An experimental study of Wiffle ball aerodynamics," Am. J. Phys. **75**, 1099-1105 (2007).

17) H. Rouse, *Elementary Mechanics of Fluids*, Dover Publications, New York, 1946 pp. 275-77.

18) G. R. Fowles and G. L. Cassiday, "*Analytical Mechanics* 5$^{th}$ ed, Thompson Brooks/Cole Belmont, CA, 2005, p. 69.

19) W. S. Porter and R. E. Tremblay, "A medieval example of energy conservation," Phys. Teach. **32**, 476 – 477 (1994).

20) R. Cross, "Physics of overarm throwing," Am. J. Phys. **72**, 305-313 (2004).

21) The following is a link for general information on the sport of jai-alai: <http://www.jai-alai.info/>

22) The compiler, example programs, and full documentation can be found at <http://vpython.org/>

23) J. O'Connell, "Dynamics of a medieval missile launcher: the trebuchet," Phys. Teach. **39**, 471 – 473 (2001).

Table I. The optimized parameters and the resulting ranges of the seven siege engine designs considered.

| Engine | Optimized Quantities | Range (m) |
|---|---|---|
| Tosser | θT = 57° | 5.9 |
| Slinger | SL = 1.3 m | 32 |
| Roller | RL = 0.63 m | 10 |
| Hinged Roller | RH = 0.27 m, ψ = 38.9°, A = 0.06 m | 18 |
| Torsion Tosser | θT = 68° | 5.4 |
| Torsion Slinger | SL = 0.8 m | 20 |
| Torsion Roller | RL = 0.52 m | 8.9 |

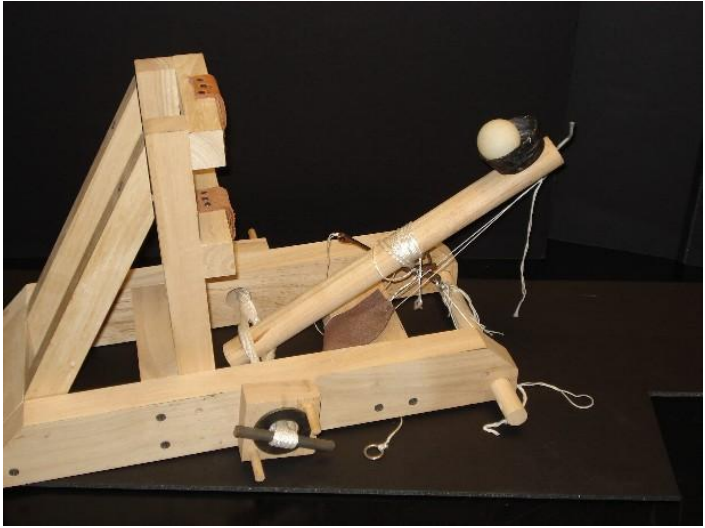

Fig. 1. A torsion powered spoon design, often referred to as an onager. When loosed, the ball will leave the spoon before the spar hits the cross beam. The kit shown is sold as a torsion powered sling launcher, the cup was attached for illustration purposes only.

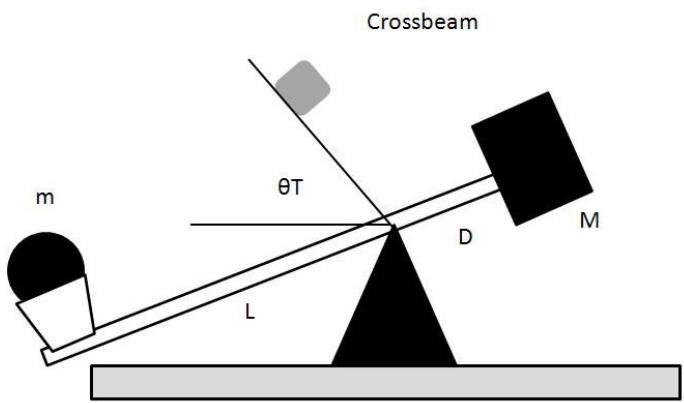

Fig. 2. In the Tosser, the ball is projected from the cup, at the point the spar impacts with the crossbeam. The range is determined by the location of the crossbeam. Notice the similarities with fig. 1.

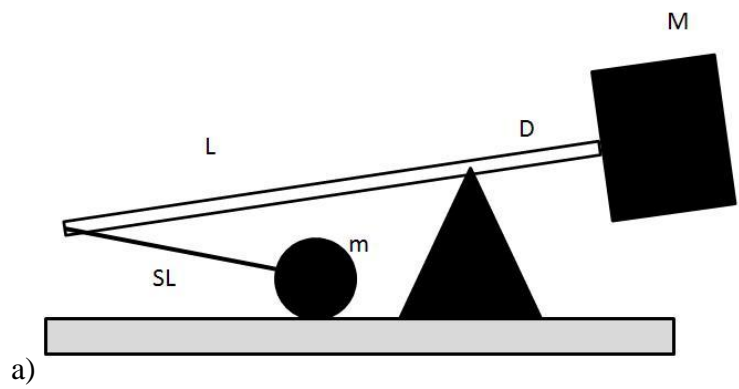

a)

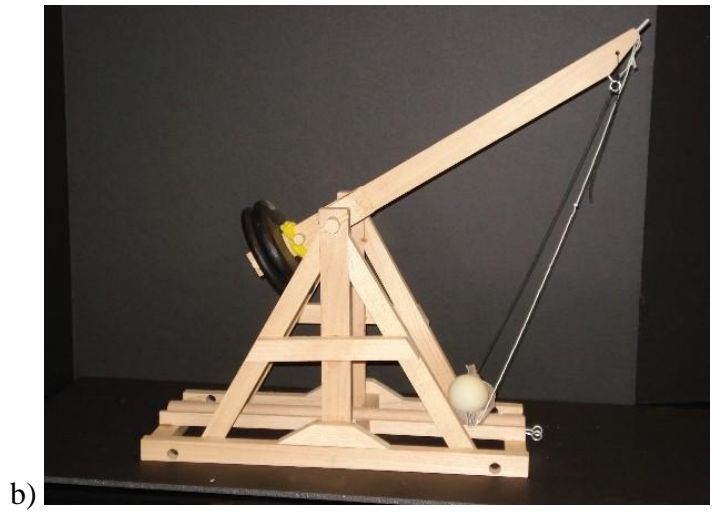

b)

Figure 3. The Slinger design. Illustrated as a) schematic, and b) photograph of commercial kit constructed by the authors, and armed with a ping-pong ball.

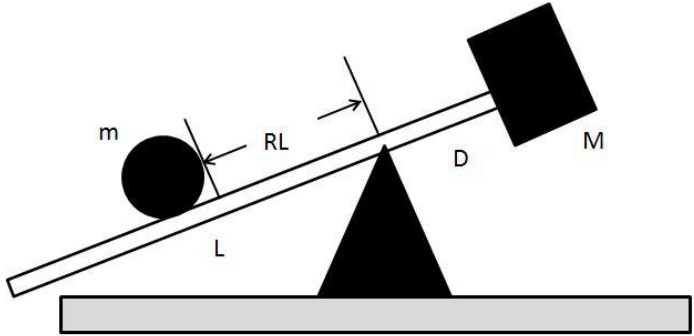

Fig. 4. The Roller design. The ball is "released" to begin free flight when it reaches the end of the spar. Note, in comparison with the Tosser of Fig. 2, a crossbeam is not necessary. After the ball is released, the crew allows the spar oscillations, to decrease in amplitude until it is safe to begin the reloading process.

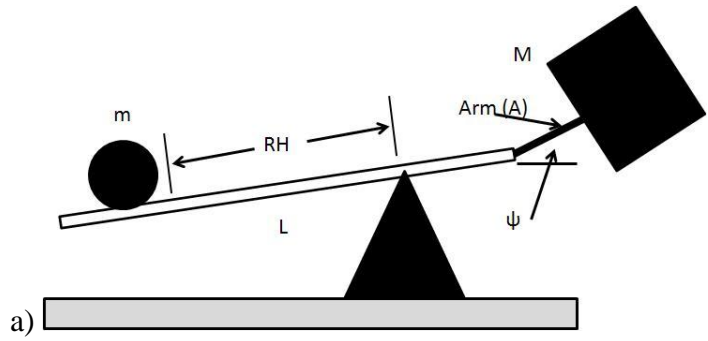

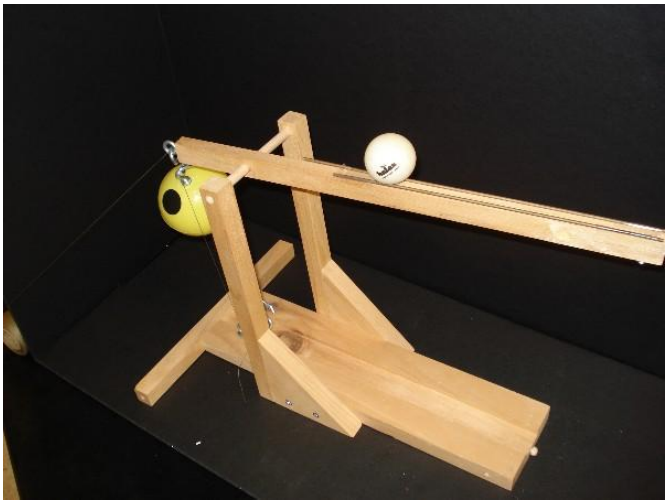

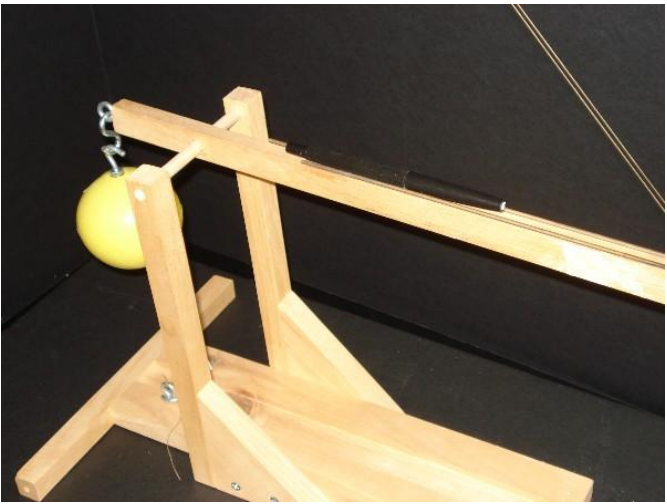

Fig. 5. The Hinged Roller (a, b) and Hinged Slider variant (c). The support for the counterweight (M) is hinged. The total length of the support is the same as in the Roller (0.3 m),

but a portion of the support is hinged with a length of A (A < D).  Illustrated as a) schematic; b) is a photograph of a commercial trebuchet kit constructed by the authors and modified by the removal of the sling and the addition of the guide rails.  In c) the Hinged Slider variant is shown loaded with a pen between the guide rails.  Caution should be exercised with the Slider variant.  Use only bolts with blunt ends.